\newcommand{\PRL}[3]{Phys.\ Rev.\ Lett.\ {\bf #1},\ #2 (#3)}
\newcommand{\RMP}[3]{Rev.\ Mod.\ Phys.\ {\bf #1},\ #2 (#3)}

\newcommand{\SC}[3]{Science\ {\bf #1},\ #2 (#3)}

\newcommand{\PRA}[3]{Phys.\ Rev.\ A\ {\bf #1},\ #2 (#3)}

\newcommand{\appsection}[1]{\let\oldthesection\thesection
  \renewcommand{\thesection}{ \oldthesection}
  \section{#1}\let\thesection\oldthesection}

\documentclass[twocolumn,preprintnumbers,amsmath,amsfonts,amssymb,notitlepage,showpacs,pra]{revtex4-1}
\usepackage{geometry}
\date{\today}
\DeclareMathAlphabet{\mathpzc}{OT1}{pzc}{m}{it}
\usepackage[utf8x]{inputenc}
\usepackage{amsmath}
\usepackage{amsfonts}
\usepackage{amssymb}
\usepackage{graphicx}
\usepackage{nicefrac}
\usepackage{amsbsy}
\usepackage{float}
\usepackage{epstopdf}
\usepackage{dsfont}
\usepackage{siunitx}

\def \be{\begin{equation}}
\def \ee{\end{equation}}
\def \ba{\begin{array}}
\def \ea{\end{array}}
\def \bea{\begin{eqnarray}}
\def \eea{\end{eqnarray}}
\def \myperp{\! \perp}

\begin{document}
\title{Transverse collisional instabilities of a Bose-Einstein condensate in a driven one-dimensional lattice}
\author{Sayan Choudhury}
\email{sc2385@cornell.edu}
\author{Erich J Mueller}
\email{em256@cornell.edu}
\affiliation{Laboratory of Atomic and Solid State Physics, Cornell University, Ithaca, New York}
\begin{abstract}
Motivated by recent experiments, we analyze the stability of a three-dimensional Bose-Einstein condensate  (BEC) loaded in a periodically driven one-dimensional optical lattice. Such periodically driven systems do not have a thermodynamic ground state, but may have a long-lived steady state which is an eigenstate of a ``Floquet Hamiltonian". We explore collisional instabilities of the Floquet ground state which transfer energy into the transverse modes. We calculate decay rates, finding that the lifetime scales as the inverse square of the scattering length and inverse of the peak three-dimensional density. These rates can be controlled by adding additional transverse potentials.
\end{abstract}

\maketitle
\section{Introduction}
In recent years, rapid progress has been made in quantum simulation, whereby one engineers a quantum system to study important phenomena experimentally \cite{NoriRMP, Ciracnatpcomm,Blochnatpcomm,Blattnatpcomm,Guziknatpcomm}. Periodically driven quantum systems (Floquet systems) are a particularly versatile platform for such simulations \cite{MoessnerFTIReview,HolthausFloquetReview} and have already been used to explore a variety of rich physics. This program has been particularly successful in cold atoms, where periodic driving has been integral to studying models of classical frustrated magnetism, and models of topological matter \cite{FerrariACTunnelingNatPhys2009,SengstockFrustratedScience2011,SengstockIsing2013,SengstockAbelianGaugePRL2012,EsslingerHaldaneFloquet, Blochbandtopology, Ketterlespinorbit, BlochHarper, KetterleHarper,TinoNJP2010,TinoPRL2008,MiyakeThesis,ChinFloquet2013,ChinFloquet2014}. These periodically driven systems have seen extensive theoretical modelling \cite{SengstockNonAbelianGaugePRL2012, ZhaiFTIarxiv, DemlerMajoranaPRL2011,DemlerFloquetTransport, OhMajoranPRB2013, BaurCooperFloquet2014,EckardtACEPL,DemlerFloquetAnamolous, ZhaoPRLAnamolous,MuellerFloquetAnamolous,creffieldsmf,GoldmanDalibardprx,PolkovnikovAnnals2013Floquet, RigolFloqetArxiv2014Floquet, EckardtPRLBoseSelection2013, DasMoessnerPRL,DasMoessner2014-2,SenguptaSensarma2013,ChandranAbaninMBLfloquet,chinz2theoryprl,polkovnikov2014floquet,EckardtFloquetPRL2005,Gomez-LeonFloquetDimensionPRL,NeupertFloquetPRL2014,TorresPRBGraphene2012,TorresPRBGraphene2014,GalitskiFTINatPhys2011,GaliskiFTIPRB2013,PodolskyFTI2013,BarangerFloquetMajorana,KunduSeradjehMajoranaPRL,Cooperarxiv2014,Cooperdalibardarxiv2014,Demlerarxiv2015}. Some of these experiments have experienced unexpected heating \cite{KetterleHarper}. In an earlier paper, we began addressing the sources of this heating by studying collisions within a one-dimensional BEC in a shaken optical lattice \cite{ChoudhuryMuellerPRA2014}. We found that in the presence of strong transverse confinement, interactions can drive instabilities but that there were large parameter ranges where the system was stable. Here we extend that work to the regime where there is no transverse confinement. The additional decay channels generally lead to more dissipation and diffusive dynamics.  \\

\begin{figure}
\begin{center}
\includegraphics[scale=0.4]{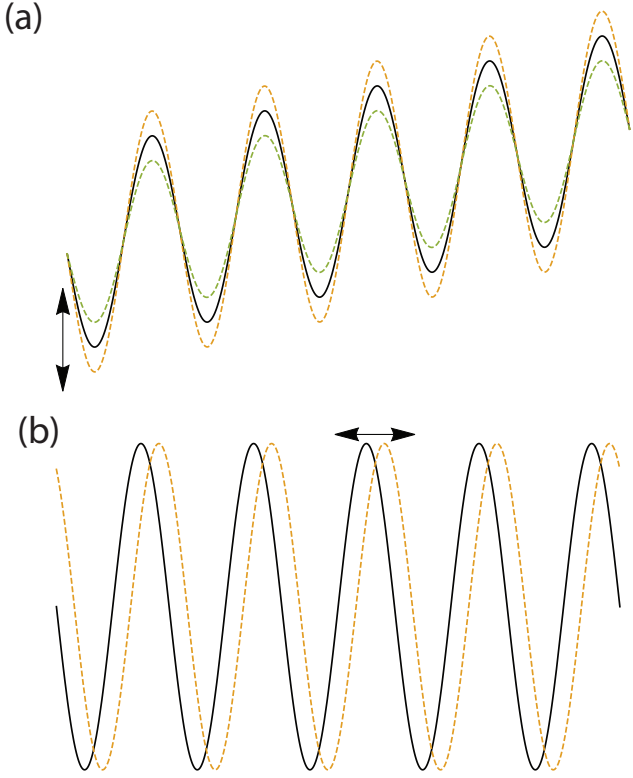}
\caption{(Color Online) The two protocols of lattice driving (a) An amplitude modulated tilted lattice and (b) A shaken lattice}
\label{fig1}
\end{center}
\end{figure}

In this paper, we consider two paradigmatic examples of Floquet systems in which a three dimensional BEC is loaded into an a modulated one-dimensional lattice. The difference lies in the nature of the drive: We consider (a) amplitude modulation of lattice depth (similar to the setup in Refs. \cite{TinoNJP2010,TinoPRL2008,MiyakeThesis}) and (b) lattice shaking (similar to the setup in Ref. \cite{ChinFloquet2013,ChinFloquet2014}). These two protocols are illustrated schematically in Fig. \ref{fig1}. We solve the Schr\"{o}dinger equation for both systems and treat the inter-atomic interactions perturbatively. Our analysis is along the lines of Ref. \cite{ChoudhuryMuellerPRA2014} where we used Fermi's golden rule to study the tight confinement limit. This kinetic approach can be contrasted with quantum coherent arguments such as those used  by Creffield in Ref. \cite{CreffieldPRA2008}. Creffield used the Bogoliubov equations to look at  a dynamical instability of a BEC in a shaken one dimensional optical lattice. These decay channels are important when the interactions are strong. We consider a different limit: for most recent experiments, the interaction strengths are too low for the interaction-driven modification of the dispersion to be relevant, rather the physics is dominated by the energy and momentum conserving scattering processes which are accounted for through our kinetic equations. In a field-theoretic formulation this corresponds to only keeping the imaginary part of the self-energy.\\

In section II, we analyze  the stability of a BEC in an amplitude modulated tilted optical lattice. A similar analysis can be used for Raman-driven lattices, such as those used to realize the Harper Hamiltonian \cite{KetterleHarper, MiyakeThesis}. It also applies to the study of density induced tunnelling \cite{Sengstockdit} and is related to earlier studies of Bloch oscillations \cite{ArimondoBloch2001}. In section III, we study the stability of a BEC loaded in a shaken optical lattice. This system can be mapped onto a classical spin model which exhibits a paramagnetic-ferromagnetic phase transition as well as a roton-maxon excitation spectrum \cite{ChinFloquet2013,ChinFloquet2014}.  In both section II and section III, we obtain analytical results for the lifetime of the BEC. Finally, in section IV, we discuss the general form of the dissipation rate in driven systems.\\
\section{Amplitude Modulated Lattice}
In this section, we consider a BEC in a deep tilted one dimensional optical lattice. Adjacent sites are offset by an energy $\Delta \gg J$, suppressing tunneling ($J$ being the nearest neighbor tunnelling matrix element). There is no transverse confinement, yielding a one-dimensional array of pancakes. The lattice depth is then modulated at a frequency $\omega$($\approx \Delta$) so that tunnelling is restored between the pancakes. The Hamiltonian describing this system is :
\begin{eqnarray}\label{model}
H&=&\int d^2 r_{\myperp} \sum_j -\left(J+2 \Omega \cos(\omega t)\right) \left(a_{j+1}^\dagger a_j+a_j^\dagger a_{j+1}\right)\nonumber\\
&+& \Delta j a_j^\dagger a_j  + \frac{\overline{g}}{2} a_j^\dagger a_j^\dagger a_j a_j +\frac{\hbar^2}{2m} \nabla_{\myperp} a_j^{\dagger} \nabla_{\myperp} a_j,
\end{eqnarray}
The constant $\Omega$ parameterizes the modulation of the hopping matrix element. The transverse spatial components are  suppressed : $a_j = a_j (r_{\myperp})$ where $r_{\myperp} = (x,y)$ and $\nabla_{\myperp} = \hat{x} \partial_x + \hat{y} \partial_y$. The coupling constant is
\begin{eqnarray}
\overline{g}&=&\frac{4\pi \hbar^2 a_s}{m} \int\! dz\, \phi(z)^4 \nonumber \\
&=&\frac{4\pi \hbar^2 a_s}{m d} 
\label{wan}
\end{eqnarray}
where $\phi(z)$ is the Wannier wavefunction in the $z$ direction, normalized so that $\int |\phi|^2 dz=1$ and $a$ is the lattice spacing. This equation defines $d$, the size of the Wannier state and is valid if $d \gg a_s$ \cite{HazzardMuellerPRA2010}.\\

Depending on how one sets up the problem the $\phi(z)$ used in Eq.(\ref{wan}) will be either the Wannier states of the static lattice, some time average of the instantaneous eigenstates or even some time-dependent function which yields an oscillating $\overline{g}$. The distinction will be important if the drive frequency is resonant with a band changing collision or if the modulation amplitude is large. Similarly, the relationship between $J, \Omega$ and the lattice parameters may be renormalised by large amplitude driving and the time-dependence of the parameters may not be sinusoidal. For most present experiments, where the amplitude of oscillations is small, these effects can be ignored. \\

As in \cite{KolovskyPRA2009}, we now perform a gauge transformation to replace the tilt with a time dependent phase :
\begin{equation}
a_j=b_j e^{-i \Delta j t}.
\end{equation}
The operators $b_j$ will evolve with a new Hamiltonian $H^\prime$, chosen so that
\begin{equation}
i\partial_t b_j = [b_j,H^\prime].
\end{equation}
Specializing to the resonant case $\omega=\Delta$, we Fourier transform this equation yielding
\be\label{rf}
H^\prime = \sum_k \epsilon_{\bf k}(t) b_{\bf k}^\dagger b_{\bf k} + \frac{g}{2 V}\sum_{\bf k_1,k_2,k_3}b_{\bf k_1}^\dagger b_{\bf k_2}^\dagger b_{\bf k_3} b_{\bf k_4},
\ee
where ${\bf k_4 = k_1+ k_2 - k_3}$, ${\bf k} = \{k_z, k_{\perp}\}$ and $g=\overline{g}a$, where $a$ is the lattice spacing. The instantaneous single-particle dispersion is given by:
\begin{eqnarray}
\epsilon_{\bf k}(t)&=& -2\Omega \cos(k_z) -2\Omega \cos(k_z-2 \Delta t) \nonumber \\
&-& 2 J \cos(k_z-\Delta t) + \frac{\hbar^2 k_{\myperp}^2}{2m} 
\end{eqnarray}
where $V$ is the system volume and $b_{\bf k} = \sum_j b_j \exp(i {\bf k} j)$. The best interpretation of this dispersion comes from looking at the group velocity of a wave-packet, $\partial \epsilon/\partial k$. There is a drift term, $v_d = \partial \epsilon/\partial k_z = 2 \Omega \sin(k_z)$ and an oscillating part $v_m = \partial \epsilon/\partial k_z = -4 \Omega \Delta \sin(k_z-2 \Delta t) - 2 J \sin(k_z-\Delta t) $  which is analogous to micro motion in ion traps \cite{Winelandjappphys98}
\\

We wish to explore the behaviour of a condensate at ${\bf k}=0$. To this end, we break our Hamiltonian into three terms $H^\prime=H_0+H_1+H_2$, 
\begin{eqnarray}
H_0&=&  \sum_{\bf k} \epsilon_{\bf k}(t) b_{\bf k}^\dagger b_{\bf k} +\frac{g}{2 V} b_0^\dagger b_0^\dagger b_0 b_0  + \frac{2 g}{V}\sum_{\bf k\neq 0} b_0^\dagger b_{\bf k}^\dagger b_{\bf k} b_0, \nonumber\\
 \\
H_1&=&  \alpha \frac{g}{2 V}\sum_{\bf k\neq 0} b_{\bf -k}^\dagger b_{\bf k}^\dagger b_0 b_0+{\rm H. C.},\\
H_2 &=& H-H_1-H_0
\end{eqnarray}
where $\alpha=1$ is a formal parameter we will use for perturbation theory. As $\alpha$ is accompanied by a factor of the interaction strength $g N/V$, this expansion is equivalent to perturbation theory in $g$. Here $H_0$ contains the single-particle physics and the Hartree-Fock terms, $H_1$ contains interaction terms corresponding to atoms scattering from the condensate to finite momentum states and $H_2$ contains terms where a condensed and a non-condensed atom scatter or two non-condensed atoms scatter. $H_2$ does not contribute at lowest order in perturbation theory, as there are initially no non-condensed atoms. \\

We will imagine that at time $t=0$ we are in the state
\begin{equation}
|0\rangle =\frac{\left(b_0^\dagger\right)^N}{\sqrt{N!}} |{\rm vac}\rangle,
\end{equation}
which is an eigenstate of $H_0$. We will perturbatively calculate how $|\psi(t)\rangle$ evolves. To lowest order,
\begin{eqnarray}
|\psi(t)\rangle
&=& e^{-i \frac {E_0 t}{\hbar}}\left[ |0\rangle+\sum_{\bf k} c_{\bf k}(t) |{\bf k}\rangle +\cdots\right]\
\end{eqnarray}
where the state $\vert \bf k \rangle$ is given by :
\begin{equation}
\vert {\bf k}\rangle = b_{\bf k}^\dagger b_{\bf -k}^\dagger \frac{\left(b_0^\dagger\right)^{N-2}}{\sqrt{(N-2)!}} |{\rm vac}\rangle.
\end{equation}
and the coefficient is 
\begin{equation}\label{sol}
c_{\bf k}(t) =\frac{\Lambda_k}{i \hbar}\int_0^t\!d\tau\,\exp\left[
-i \int_\tau^t 2 \frac{E_k(s)}{\hbar} \,ds
\right].
\end{equation}
whose amplitude is given by 
\be
\Lambda_k = \langle {\bf k} | H_1 |0\rangle/\alpha = \frac{g n}{2} 
\ee
In Eq.(\ref{sol}), the (Hartee-Fock) excitation energy is 
\begin{equation}
E_k(t) = \epsilon_{\bf k}(t)+g n-\epsilon_0(t).
\end{equation}

Performing the integral in the exponent yields
\begin{eqnarray}
\int_\tau^t E_k(s)\,ds&=& E^{(0)}_k \times(t-\tau) \nonumber \\
&+& \frac{\Omega}{\Delta} \left(\sin(k_z-2\Delta \tau) - \sin(k_z-2\Delta t) \right) \nonumber \\
&+& \frac{2 J}{\Delta} \left( \sin(k_z-\Delta \tau) - \sin(k_z-\Delta t)\right) \nonumber\\
\end{eqnarray}
where the ``effective dispersion" is
\begin{eqnarray}
E_k^{(0)}&=&2\Omega [1-\cos(k_z)] + g n +\frac{k_{\myperp}^2}{2 m}.
\label{jeff}
\end{eqnarray}
\\

This energy corresponds to the spectrum one would obtain from Floquet theory. It takes the form of a tight-binding model along z with a nearest-neighbor hopping of strength $\Omega$. The resonant modulation has restored hopping. We now expand Eq.~(\ref{sol}) in powers of
$J/\Delta$ and $\Omega/\Delta$.  Neglecting off-resonant terms and making the standard approximation $\sin^2(xt) /(xt)^2 \approx 2 \pi t \delta(x)$, finding

\begin{eqnarray}
|c_{\bf k}|^2 &\approx& \frac{|\Lambda_k|^2}{\hbar} \frac{\Omega^2}{\Delta^2} t\, 2\pi \delta(E^{(0)}_k-\Delta) \nonumber \\
&+& \frac{|\Lambda_k|^2}{\hbar} \frac{4 J^2}{\Delta^2} t \,2\pi \delta(E^{(0)}_k-\Delta/2),
\end{eqnarray}
which is analogous to Fermi's golden rule. The result can also be derived using the formulation in Ref. \cite{floquetunitary}. The first term proportional to $\Omega^2$ is naturally interpreted as coming from a pair of particles absorbing a lattice vibration. The second term involves one particle ``hopping downhill" with the potential energy converted to transverse motion.\\

We now calculate the total rate of scattering out of the condensate.  The relevant timescale is
\begin{eqnarray}
\frac{1}{\tau}&=&\frac{1}{N_0}\partial_t N_0=\frac{2}{N}\partial_t \sum_k |c_{\bf k}|^2 \nonumber\\
&=& \frac{1}{\tau_2}+\frac{1}{\tau_1} \nonumber \\
\frac{1}{\tau_2}&=& \frac{2 |\Lambda_k|^2}{N \hbar} \frac{\Omega^2}{\Delta^2}\sum_k 2\pi \delta(E^{(0)}_k-\Delta)\\
\frac{1}{\tau_1}&=&\frac{2 |\Lambda_k|^2}{N \hbar} \frac{4 J^2}{\Delta^2}\sum_k 2\pi \delta(E^{(0)}_k-\Delta/2).
\end{eqnarray}
The sums over $k$ are straightforward.  We first note that that because $\Omega$ is small, the dependence of $E_k^{(0)}$ on $k_z$ is weak, and can be neglected.  Thus
the sum over $k$ just yields a constant
\begin{eqnarray}
\rho(\nu)&=&\sum_k 2\pi \delta(E^{(0)}_k-\nu) \nonumber \\
&\approx& \frac{V}{a}\int \frac{d^2 k_{\myperp}}{(2\pi)^2} 2\pi\delta\left(\frac{k_{\myperp}^2}{2m}+g n-\nu\right) \nonumber \\
&=& \frac{V m}{a}.
\end{eqnarray}
Putting in the factors of $\hbar$ the total rate of scattering out of the condensate is
\begin{eqnarray}
\frac{1}{\tau}&=&\frac{g^2 n m}{2a \hbar^3} \frac{\Omega^2+4 J^2}{\Delta^2} \nonumber \\
&=& gn \frac{2\pi a_s}{\hbar d}\frac{\Omega^2+4 J^2}{\Delta^2}
\label{t1}
\end{eqnarray}
Some typical numbers are $gn/h\sim 300 $Hz$, \Omega\sim 40 $Hz$, J\sim 5$Hz$, \Delta\sim 1 $kHz and $d\sim 75$nm.  For $^{87} {\rm Rb}$, the scattering length is $a_s\sim 5$nm.  Thus the lifetime of the  BEC is about $750$ms.

\section{Shaken Lattice}
\begin{figure}
\begin{center}
\includegraphics[scale=0.8]{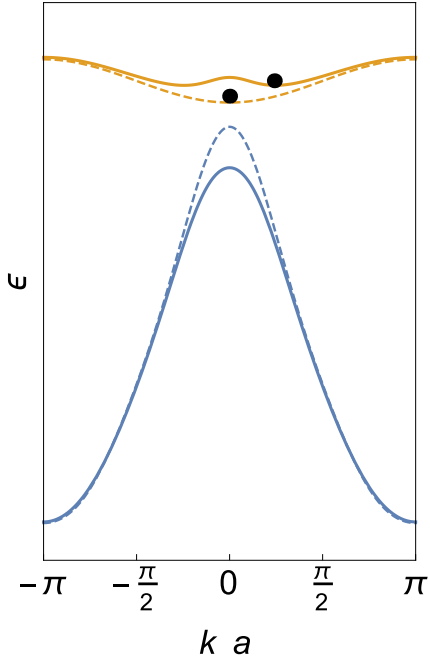}
\caption{(Color Online) Schematic showing first (top) and second (bottom) Floquet quasi-energy bands of an optical lattice: $\epsilon$ is the single-particle energy, $k$ is the quasi-momentum and $a$ is the lattice spacing. Since Floquet energies are only defined modulo the shaking quanta $\hbar \omega$,  the energy of the second band has been shifted down by $\hbar \omega$. Alternatively, this shift can be interpreted  as working in a dressed basis, where the energy includes a contribution from the phonons. The mixing between the bands depends on the shaking amplitude. Dashed curves correspond to weak shaking, where the first band has its minimum at $k = 0$. Solid curves correspond to strong shaking, where there are two minima at $k = \pm k_0 \ne 0$.}
\label{fig2}
\end{center}
\end{figure} 

In this section, we look at the stability of a three-dimensional BEC loaded into a shaken one-dimensional optical lattice. We considered the strictly one-dimensional version in Ref. \cite{ChoudhuryMuellerPRA2014}. We are motivated by the set-up in Ref. \cite{ChinFloquet2014} where Ha {\it et al.} load a three-dimensional BEC of $^{133} {\rm Cs}$ atoms in a one-dimensional lattice and then shake the lattice at a frequency resonant with the zero-energy bandgap of the first two bands. This results in a strong mixing of the first two bands (schematically illustrated in Fig. \ref{fig2}). For our analysis, we we label the Bloch band connected adiabatically to the first Bloch band in the limit of zero shaking as the ground band. As is evident from Fig. \ref{fig2}, due to level repulsion between the Bloch bands, the ground band exhibits a bifurcation from having one minimum at $\{{\bf k}=0 \}$ to two minima at $\{{\bf k_{\myperp} } = 0, k = k_0 \ne 0\}$. This is analogous to the paramagnetic-ferromagnetic phase transition in Landau theory for classical spin models. In the paramagnetic regime the bosons always condense at ${\bf k} = 0$, while in the ferromagnetic regime, the bosons condense at some finite momentum $\{{\bf k_{\myperp}} = 0, k \ne 0\}$. Here, we first perturbatively analyze the stability of a BEC against collisions in the limit of weak forcing amplitude. This gives an intuitive picture about how the scattering rate varies with amplitude. We then numerically calculate collision rates for larger shaking amplitudes spanning the experimentally interesting critical region. We find that the linearised theory over-estimates the damping, but gives the correct order of magnitude.\\

\subsection{Model}
In the frame co-moving with the lattice, the tight-binding Hamiltonian describing the system can be written as $H_0 (t) + H_{\rm int}$:
\bea
H_0(t) &=& \int d^2 r_{\myperp} \sum_{ij} \left(-t_{ij} ^{(1)} a_{i}^{\dagger}a_{j} +  t_{ij} ^{(2)}  b_{i}^{\dagger}b_{j} + h.c.\right) \nonumber \\
&+&  \sum_{j} F \cos(\omega t) \left(z_j \left(a_{j}^{\dagger} a_{j} + b_{j}^{\dagger} b_{j}\right) + \chi_j a_j^{\dagger} b_j + \chi_j ^{*} b_j^{\dagger} a_j \right) \nonumber\\
&+& \frac{\hbar^2}{2m} \left(\nabla_{\myperp} a_j^{\dagger} \nabla_{\myperp} a_j + \nabla_{\myperp} b_j^{\dagger} \nabla_{\myperp} b_j \right)\\
H_{\rm int} &=& \int d^2 r_{\myperp} \sum_i \frac{\overline{g}_1}{2} a_i ^{\dagger} a_i ^{\dagger} a_i a_i + \frac{\overline{g}_2}{2} b_i ^{\dagger} b_i ^{\dagger} b_i b_i  \nonumber \\
&+&  2 \overline{g}_{12} a_i ^{\dagger} b_i ^{\dagger} a_i b_i + H^{\prime}
\eea
where,
\bea
\chi_j  &=& \int\!dz \,\ z w_1^{*}(z-z_j) w_2(z-z_j) \nonumber \\
t_{ij} ^{(1)} &=& \int dz\,\ w_1^{*}(z-z_i)\left(\frac{-\hbar^2}{2 m} \frac{d^2}{dz^2}+ V(z)\right)w_1^{*}(z-z_j)\nonumber \\
t_{ij} ^{(2)} &=& \int dz\,\ w_2^{*}(z-z_i) \left(\frac{-\hbar^2}{2 m} \frac{d^2}{dx^2}+V(z)\right)w_2^{*}(z-z_j) \nonumber 
\eea
with $V(z) = V_0 \sin^2\left(\frac{2 \pi z}{\lambda_L}\right)$ and $H^{\prime}$ is off-resonant. It should also be noted that $\chi_j$ is independent of $j$ and so we can call it $\chi$. If necessary more bands can be included. \\

We now perform a basis rotation : $\vert\psi\rangle \rightarrow U_{c}(t) \vert \psi\rangle$ with:
\be
U_{c}(t) = \exp\left(- \frac{i}{\hbar} \int_{0}^{t} \sum_{j}z_ j F_0\cos(\omega t) (a_{j}^{\dagger} a_{j} + b_{j}^{\dagger} b_{j}) \right)
\label{unitary}
\ee
Under this unitary transformation, the Hamiltonian becomes:
\bea
H_0^{\prime} (t) &=& U_{c}H_0(t)U_{c}^{-1} - i \hbar U_{c}\partial_t U_{c}^{-1} \nonumber  \\
&=& \sum_{ij} \left(-J_{ij} ^{(1)} (t) a_{i}^{\dagger}a_{j} +  J_{ij} ^{(2)}(t)  b_{i}^{\dagger}b_{j} + h.c.\right) \nonumber\\
&+&\sum_j F \cos(\omega t)  \left(\chi a_{j}^{\dagger} b_{j} +\chi ^{*} b_{j}^{\dagger} a_{j} \right)  + \sum_{k_{\myperp}}    \frac{\hbar^2 k_{\myperp}^2}{2m}\nonumber \\
&=& \sum_k \sum_m  \cos(m k a)\left(-J_{m} ^{(1)} (t) a_{\bf k}^{\dagger}a_{\bf k} -J_{m} ^{(2)} (t) b_{\bf k}^{\dagger}b_{\bf k}\right)\nonumber \\
&+& \sum_k F_0\cos(\omega t)  \left(\chi a_{\bf k}^{\dagger} b_{\bf k} +\chi ^{*} b_{\bf k}^{\dagger} a_{\bf k} \right)  + \sum_{k_{\myperp}}    \frac{\hbar^2 k_{\myperp}^2}{2m}\nonumber \\
\eea
where,
\bea
J_{ij}^{\sigma} (t) &=& t_{ij}^{\sigma} \exp(-i F_0 \frac{\cos(\omega t)}{\hbar \omega} (z_i-z_j)) \nonumber\\
&=& t_{ij}^{\sigma} \exp(-i F_0 \frac{\cos(\omega t)}{\hbar \omega} a (i-j)) ,
\label{rwa1}
\eea
$a = \lambda_L/2$ is the lattice spacing and $\chi=\chi^{*}$ for a suitable choice of phase for $a_k$ and $b_k$.\\

Thus, in the limit of $F/(\hbar \omega) \ll 1$, the Hamiltonian describing the system is : $H = H_{\rm sp} + H_{\rm int}$, where
\bea
H_{\rm sp} &=&   \sum_{\bf k} \epsilon^{(1)}_{\bf k} a_{\bf k}^{\dagger} a_{\bf k} + \epsilon^{(2)}_{\bf k} b_{\bf k}^{\dagger} b_{\bf k} +\chi F \cos(\omega t) \left (a_{\bf k}^{\dagger}b_{\bf k} + b_{\bf k}^{\dagger}a_{\bf k} \right)\nonumber\\
\\
H_{\rm int} &=& \int d^2 r_{\myperp} \sum_i \frac{\overline{g}_1}{2 } a_i ^{\dagger} a_i ^{\dagger} a_i a_i + \frac{\overline{g}_2}{2} b_i ^{\dagger} b_i ^{\dagger} b_i b_i  \nonumber \\
&+& 2 \overline{g}_{12} a_i ^{\dagger} b_i ^{\dagger} a_i b_i + H^{\prime}
\eea
Here, $\epsilon^{(1)}_{\bf k} (\epsilon^{(2)}_{\bf k})$ is the dispersion of the first (second) band and $a_{\bf k} (b_{\bf k})$ is the annihilation operator for particles in the first (second band).\\

We make the transformation $b_{\bf k} \rightarrow \exp(- i \omega t) b_{\bf k}$ and discard far off-resonant terms (making the rotating wave approximation) to simplify the single-particle terms :
\bea
H_{\rm RWA} ^{(\rm sp)} &=& \sum_{\bf k} \epsilon^{(1)}_{\bf k} a_{\bf k} ^{\dagger} a_{\bf k} + \epsilon^{(2)}_{\bf k}  b_{\bf k} ^{\dagger} b_{\bf k} \nonumber \\
&+& \chi F\left(a_{\bf k} ^{\dagger} b_{\bf k} +  b_{\bf k} ^{\dagger} a_{\bf k} \right),
\eea
Here ${\bf k} = \{k, {\bf k_{\myperp}}\}$, $\epsilon^{(1)}_{\bf k} = \epsilon^{(1)}_{k} + (\hbar k_{\myperp})^2/(2m) $, $\epsilon^{(2)}_{k} = \epsilon^{(2)}_{\bf k} + (\hbar k_{\myperp})^2/(2m) - \hbar \omega$.  We diagonalise this quadratic form writing 
\be
H_{\rm RWA} ^{(\rm sp)} = \sum_{\bf k} \overline{\epsilon}^{(1)}_{\bf k} \overline{a}_{\bf k} ^{\dagger} \overline{a}_{\bf k} + \overline{\epsilon}^{(2)}_{\bf k} \overline{b}_{\bf k} ^{\dagger}  \overline{b}_{\bf k}
\label{hamrwa}
\ee
The dressed dispersions $\overline{\epsilon}^{(1)}_{\bf k}$ and $\overline{\epsilon}^{(2)}_{\bf k}$ are shown as solid lines in Fig.(\ref{fig2}). The bare dispersions $\epsilon^{(1)}_{\bf k}$ and $\epsilon^{(2)}_{\bf k}$ are shown as dashed lines. We treat $H_{\rm RWA} ^{(\rm sp)}$ both perturbatively and non-perturbatively to obtain scattering rates in the next two subsections.

\subsection{Perturbation Theory}
For small forcing amplitudes, we gain insight by a perturbative expansion in F. To linear order in F, the dressed operators are
\bea
\overline{a}_{\bf k} ^{\dagger}  &=& a_{\bf k} ^{\dagger} - (\chi F)/(\epsilon^{(2)}_{\bf k} - \epsilon^{(1)}_{\bf k} ) b_{\bf k} ^{\dagger} \\
\overline{b}_{\bf k} ^{\dagger}  &=& b_{\bf k} ^{\dagger} + (\chi F)/(\epsilon^{(2)}_{\bf k} - \epsilon^{(1)}_{\bf k}) a_{\bf k} ^{\dagger} 
\eea

Because we have made the rotating wave approximation, we have a time-independent problem and can simply apply Fermi's Golden Rule. The standard procedure yields a scattering rate:
\bea
\frac{d N}{dt} &=& \int \frac{dk}{2 \pi} \int \frac{d^2 k_{\myperp}}{(2 \pi)^2}\vert \langle \psi_f \vert H_{\rm int} \vert \psi_i \rangle\vert ^2  \sigma \\
\sigma &=& \frac{2 \pi}{\hbar} \delta(\overline{\epsilon}^{(1)}_{\bf k} + \overline{\epsilon}^{(2)}_{\bf k} + \frac{(\hbar k_{\myperp})^2}{m} - 2 \overline{\epsilon}^{(1)}_{0}) \nonumber 
\eea
The initial and final states are
\bea \label{inf}
\vert \psi_i \rangle = \frac{(\overline{a}_{0} ^{\dagger})^N}{\sqrt{N !}} |0\rangle \nonumber \\
\vert \psi_f \rangle = \overline{b}_{\bf k} ^{\dagger} \overline{a}_{\bf -k} ^{\dagger} \frac{(\overline{a}_{0} ^{\dagger})^{(N-2)}}{\sqrt{N-2 !}} \vert 0\rangle
\eea
$\vert \psi_i \rangle$ represents all particles in the condensate, while $\vert \psi_f \rangle$ has one particle with momentum ${\bf k}$ in the dressed $b$ band and one with momentum ${\bf -k}$ in the ground band. \\

The transverse integrals are elementary and yield
\be
\frac{dN}{dt} = \frac{m}{2 \hbar^3} n^2 \int \frac{dk}{2 \pi} (\frac{g_1}{\Delta_k} - 2 \frac{g_{12}}{\Delta_0})^2 (\chi F)^2,
\label{scatter}
\ee
where $\Delta_k = \left(\epsilon^{(2)}_{k} - \epsilon^{(1)}_{k} \right)$, $\Delta_0 =\left( \epsilon^{(2)}_{0} - \epsilon^{(1)}_{ 0}  \right)$ and $g = \overline{g} a$. While Eq.(\ref{scatter}) can always be integrated numerically, we have found a sequence of approximations which let us analytically estimate the scattering rate. First, we approximate the Wannier functions as $w_1(x) = (\frac{1}{d_1^2 \pi})^{1/4} \exp(- x^2/2 d_1^2) $ and $w_2(x) = (\frac{1}{\pi d_1^2})^{3/4} x \exp(- x^2/2 d_1^2) $, where $d_1 =a/(\pi (V_0)^{1/4})$ ($V_0$ being the lattice depth expressed in units of $E_R$). Within this approximation, $g_1 \approx 2 g_{12}$,
where $g_1 = (4 \pi \hbar^2 a_s a)/(m d)$, $d  = d_1\sqrt{2 \pi}$ being the size of the Wannier state and $a_s$ is the scattering length . This is a good approximation as a numerical calculation using the exact Wannier states for the lattice in Ref. \cite{ChinFloquet2013,ChinFloquet2014} yields $g_1 =(1/0.41)\, g_{12}$. \\

As a second approximation, we note that except for $k$ near 0, $\Delta_k \gg \Delta_0$. The contribution of those parts to the integral in Eq.(\ref{scatter}) is small, allowing us to neglect the $k$ dependence of the integrand. Hence, we see that the rate of scattering is approximately:
\be
\frac{dN}{dt} \approx (g_1 n)^2 (\frac{\chi F}{\Delta_0})^2 \frac{V m}{2 a \hbar^3}
\label{chinscat}
\ee 
This gives the timescale for the scattering to be:
\be
\tau = \frac{N}{\frac{dN}{dt}} \approx \frac{2 \hbar^3 a}{m g_1^2 n} (\frac{\Delta_0}{\chi F})^2 .
\label{t2pert}
\ee
Stronger interactions, higher density and larger forcing amplitudes all increase the scattering rate.

\subsection{Beyond Perturbation Theory}
In this section, we extend our results to larger F. This allows us to probe the critical and ferromagnetic region. Generically, we write 
\bea
\overline{a}_{\bf k} ^{\dagger}  &=& u_{k} a_{\bf k} ^{\dagger} + v_{k} b_{\bf k} ^{\dagger} \\
\overline{b}_{\bf k} ^{\dagger}  &=& -v_{k} a_{\bf k} ^{\dagger} + u_{k} b_{\bf k} ^{\dagger} 
\eea
with $\vert u_{k} \vert^2 + \vert v_{k} \vert^2 = 1$. In particular, 
\bea
u_{k} &=& \frac{1}{\sqrt{1+\vert \gamma_k\vert^2}}  ; \,\, v_{k} = \frac{g_k}{\sqrt{1+\vert \gamma_k\vert^2}} \nonumber \\
 \frac{1}{\gamma_k} &=& \frac{\sqrt{4 F^2 \chi ^2 + \delta \epsilon_k^2}+\delta\epsilon_k}{2 \chi F} \nonumber \\
 \delta \epsilon_k &=& \epsilon _{k} ^{(1)}-\epsilon_{k} ^{(2)} \nonumber
\eea
One can invert the above relationships to obtain:\\
\bea
a_{\bf k} ^{\dagger}  &=& u_{k} \overline{a}_{\bf k} ^{\dagger} - v_{k} \overline{b}_{\bf k} ^{\dagger} \\
b_{\bf k} ^{\dagger}  &=& v_{k} \overline{a}_{\bf k} ^{\dagger} + u_{k} \overline{b}_{\bf k} ^{\dagger} 
\eea
For $F < F_c$ ($F_c$ being the critical shaking force), we use Eq. (\ref{inf}) as our initial and final states. For $F>F_c$, we use
\bea
\vert \psi_i \rangle &=& \frac{(\overline{a}_{{\bf k_0}} ^{\dagger})^N}{\sqrt{N !}} |0\rangle \nonumber \\
\vert \psi_f ^{(1)}\rangle &=& \overline{b}_{\bf k_0+k} ^{\dagger} \overline{a}_{\bf k_0-k} ^{\dagger} \frac{(\overline{a}_{\bf k_0} ^{\dagger})^{(N-2)}}{\sqrt{N-2 !}} \vert 0\rangle \nonumber \\
\vert \psi_f ^{(2)}\rangle &=& \overline{b}_{\bf k_0+k} ^{\dagger} \overline{b}_{\bf k_0-k} ^{\dagger} \frac{(\overline{a}_{0} ^{\dagger})^{(N-2)}}{\sqrt{N-2 !}} \vert 0\rangle \nonumber \\
\eea
The states are analogous to those in eq.(\ref{inf}). In particular, $\vert \psi_i \rangle$ has all particles in a finite momentum condensate (${\bf k_0} = \{k=k_0,{\bf k_{\myperp}} = 0\}$). \\

The scattering rate is then:
\bea
\frac{dN}{dt} &=& \int \frac{dk}{2 \pi} \int \frac{d^2 k_{\myperp}}{(2 \pi)^2}\vert \langle \psi_f^{(1)} \vert H_{\rm int} \vert \psi_i \rangle\vert ^2 \sigma_{12} \nonumber \\
&+& \int \frac{dk}{2 \pi} \int \frac{d^2 k_{\myperp}}{(2 \pi)^2}\vert \langle \psi_f^{(2)} \vert H_{\rm int} \vert \psi_i \rangle\vert ^2  \sigma_{22}
\eea
where
\bea
\sigma_{12} &=& \frac{2 \pi}{\hbar} \delta(\overline{\epsilon}^{(1)}_{k_0-k} + \overline{\epsilon}^{(2)}_{k_0+k} + \frac{(\hbar k_{\myperp})^2}{m} - 2 \overline{\epsilon}^{(1)}_{k_0}) \nonumber \\
\sigma_{22} &=& \frac{2 \pi}{\hbar} \delta(\overline{\epsilon}^{(2)}_{k_0-k} + \overline{\epsilon}^{(2)}_{k_0+k} + \frac{(\hbar k_{\myperp})^2}{m} - 2 \overline{\epsilon}^{(1)}_{k_0}) \nonumber 
\eea
In general $g_{12} = \alpha g_1$ and $g_2 = \beta g_1$. Approximating the Wannier functions with the harmonic oscillator wave functions would yield $\alpha= 1/2$ and $\beta = 3/4$. Rather than using this approximation, We numerically calculate the maximally localised Wannier functions for the experimental lattice depth of $V = 7 E_R$ and find that $\alpha = 0.41$ and $\beta = 0.6$.\\

Extracting the dimensional factors ,
\be
\tau = \frac{N}{\frac{dN}{dt}} = \frac{2 \hbar^3 a}{ m g_1^2 n \Gamma}
\label{t2}
\ee
where the dimensionless parameter $\Gamma$ depends on the forcing strength and can be expressed as
\begin{widetext}
\bea
\Gamma &=& \int \frac{dk}{2 \pi} (\vert - u_{k_0-k} v_{k_0+k} u_{k_0} u_{k_0} + \alpha \, u_{k_0+k} v_{k_0-k} v_{k_0} v_{ k_0} +  2\,\beta(u_{ k_0+k} u_{k_0-k} u_{k_0} v_{k_0}  - v_{k_0+k} v_{k_0-k} u_{k_0} v_{ k_0}) \vert ^2) \nonumber \\
&+& (\vert v_{ k_0-k} v_{ k_0+k} u_{ k_0} u_{ k_0} + \alpha\, u_{ k_0+k} u_{ k_0-k} v_{ k_0} v_{ k_0} - 2\, \beta(v_{ k_0+k} u_{ k_0-k} u_{ k_0} v_{ k_0}  + u_{ k_0+k} v_{ k_0-k} u_{ k_0} v_{ k_0} ) \vert ^2)
\label{scatter1}
\eea
\end{widetext}

\begin{figure}
\begin{center}
\includegraphics[scale=0.55]{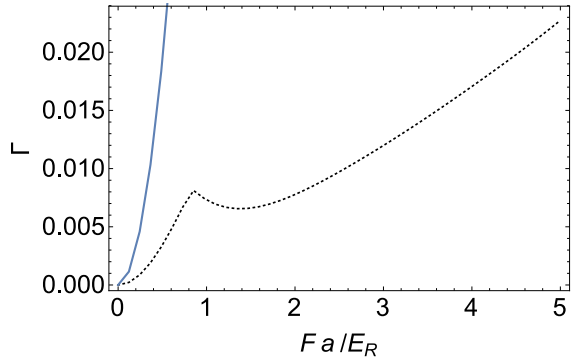}
\caption{(Color Online) Plot of dimensionless decay rate $\Gamma$ as a function of amplitude of shaking, $F$ for $\omega = 5.5 \,\ E_R/\hbar$ and $V_0 = 7.0 E_R$. The dotted line shows $\Gamma$ calculated using Eq.(\ref{scatter1}), while the thick line shows the function $ (\frac{\chi F}{\Delta_0})^2$ corresponding to the rate in Eq.(\ref{t2pert}). The kink shows the paramagnetic-ferromagnetic phase transition.}
\label{chin}
\end{center}
\end{figure} 
 
The dotted line in Fig.(\ref{chin}) shows $\Gamma$ using $\alpha = 0.41$ and $\beta = 0.6$ corresponding to a lattice depth of $V= 7 E_R$ There is a distinct kink in the $\Gamma$ vs $F$ plot which shows the paramagnetic-ferromagnetic phase transition. For all $F$, the numerical calculation gives a smaller $\Gamma$ than the perturbative estimate in Eq.(\ref{chinscat}). For the experimental lattice depths, $d \sim 100$nm, $gn/h \sim 150$Hz, $a_s \sim 1.5$nm yielding $\tau \sim 1$s which matches experimental observations \cite{ChinFloquet2013}.

\section{General Conclusions}
\subsection{Form of the scattering rate}
Generically two-particle scattering will give a rate proportional to $g^2 n$. The instabilities studied here relied upon scattering into transverse modes. These rates can be modified by tuning the density of these modes. For example, one could imagine engineering band gaps with transverse optical lattices. Note, such lattices may provide additional confinement and increase the effective $g$, inadvertently increasing the decay rate.

\subsection{Diffusive Dynamics}
The same dissipation which causes the condensate to decay can also lead to diffusive motion. Such diffusion may provide another way to study this physics. We model the kinetics by a Boltzmann equation:
\be
\frac{\partial n (z,p) }{\partial t} + v(p) \frac{\partial n (z,p) }{\partial z} = \frac{n(z,p) - (n (z)/2\pi)}{\tau}
\ee
Here $n(z,p)$ is the coarse-grained number of particles whose position along the lattice direction is $z$ and whose quasi-momentum in that direction is $p$, while $n(z) = \int dp \, n(z,p)$ is the linear density and the group velocity is $v(p) = \partial{\epsilon}/\partial{p}$. We have integrated over the transverse directions. The $\tau$ appearing here is exactly the same as in Eqs.(\ref{t1}), (\ref{t2pert}) and (\ref{t2}). The collision term takes this simple form because atoms are scattered to random values of momentum in the lattice direction after a collision. Taking the zeroth and first moments of the Boltzmann equation yields typical hydrodynamic equations
\begin{eqnarray}\
\frac{\partial n(z)}{\partial t} + \frac{\partial J}{\partial z} &=& 0 \\
 \frac{\partial J}{\partial t} + \frac{\partial}{\partial z}(\langle v^2 \rangle n(z)) &=& \frac{J}{\tau}
\end{eqnarray}
where the current $J$ is defined by $J = \int dv \, v(p) n(z,p)$. In the over damped limit, these can be rewritten as a diffusion equation with diffusion constant $D= \langle v^2\rangle \tau \propto J_{\rm eff} ^2 \tau$, where $J_{\rm eff}$ is the effective tunnelling coefficient (cf. Eq.(\ref{jeff})). Observing the diffusive motion may be one way of experimentally measuring $\tau$, complementing more direct methods \cite{SchneiderNATP2012, SchneiderPRL2013} 

\section{Summary and Outlook}
In this paper we analysed the stability of a BEC in a driven one-dimensional optical lattice with no transverse confinement. We found that due to the presence of transverse modes, the BEC would always be unstable and we calculate the decay rates. Experimentally, this instability would be manifest in many forms, including heating and diffusive dynamics. In previous work, we found that in the limit of extremely tight transverse confinement the BEC has regimes of stability. \\

Generally, experiments are neither in the tight binding limit, nor in the limit with no transverse confinement. The results in the present paper are applicable as long as the level spacing of the quantum modes in the transverse direction ($\sim 100$ Hz for the experiment in Ref.\cite{ChinFloquet2014}) are small as compared to the drive frequency $\omega$ ($\sim 7.3$ KHz for the experiment in Ref.\cite{ChinFloquet2014}). The results from \cite{ChoudhuryMuellerPRA2014} apply in the opposite limit.\\

\section*{Acknowledgements}
We thank the Ketterle group (Wolfgang Ketterle, Colin J. Kennedy, William Cody Burton and  Woo Chang Chung) and the Chin group (Cheng Chin and Logan Clark) for correspondence about their experiments. We are particularly indebted to Wolfgang Ketterle for suggesting we investigate transverse mode instabilities. We acknowledge support from ARO-MURI Non-equilibrium Many-body Dynamics grant (W911NF-14-1-0003).

\end{document}